\documentclass[a4paper,12pt,leqno]{article}

\usepackage{latexsym}
\usepackage{bm,amsfonts,amsmath,amssymb} 
\usepackage[utf8]{inputenc}
\usepackage{abstract}
\usepackage[T1]{fontenc}
\usepackage{graphicx}
\usepackage{verbatim}
\usepackage{ulem}
\usepackage[perpage]{footmisc}
\usepackage{fancybox}
\usepackage{dsfont}

\begin{document}

\newcommand{\R}{\mathbb{R}}
\newcommand{\C}{\mathbb{C}}
\newcommand{\N}{\mathbb{N}}
\newcommand{\Z}{\mathbb{Z}}
\newcommand{\Q}{\mathbb{Q}}
\newcommand{\mj}{\mathcal}

\newcommand{\fcv}{\rightharpoonup}
\newcommand{\bg}{\begin}
\newcommand{\ds}{\displaystyle}
\newcommand{\Om}{\Omega}
\newcommand{\eps}{\epsilon}
\newcommand{\Sp}{\mathbb{S}}
\newcommand{\inj}{\hookrightarrow}
\newcommand{\n}{\textbf{n}}
\newcommand{\bv}{\textbf{b}}
\newcommand{\h}{\textbf{h}}
\newcommand{\A}{\textbf{A}}
\newcommand{\B}{\textbf{B}}
\newcommand{\F}{\textbf{F}}
\newcommand{\No}{\textbf{N}}
\newcommand{\dr}{\partial}
\newcommand{\vv}{\textbf{v}}
\newcommand{\uu}{\textbf{u}}
\newcommand{\tr}{\mathrm{Tr}}
\newcommand{\su}{\mathrm{supp}}
\newcommand{\conj}{\overline}
\newcommand{\intn}{\int_{\Om}\!\!\!\!\!\!\!-}
\newcommand{\dive}{\mathrm{div}}
\newtheorem{lem}{Lemma}[section]
\newtheorem{theo}[lem]{Theorem}
\newtheorem{prop}[lem]{Proposition}
\newtheorem{sch}[lem]{Scholie}
\newtheorem{cor}[lem]{Corollary}

\def\got#1{{\bm{\mathfrak{#1}}}}

\newenvironment{preuve}
{\noindent{\textbf{Proof.}}\\\rm\noindent}
{\bg{flushright}\tiny $\blacksquare$\end{flushright}}

\newenvironment{rem}
{\noindent\addtocounter{lem}{1}
{\textbf{Remark \thelem.}}\\\noindent\rm}
{\bg{flushright}\tiny $\blacksquare$\end{flushright}}

\renewcommand{\theequation}{\thesection.\arabic{equation}}
\title{Uniform spectral estimates for families of Schrödinger operators with magnetic field of constant intensity and applications}
\author{Nicolas Raymond}
\maketitle{}

\bg{abstract}
The aim of this paper is to establish uniform estimates of the spectrum's bottom of the Neumann realization of $(i\nabla+q\A)^2$ on a bounded open set $\Om$ with smooth boundary when $|\nabla\times\A|=1$ and $q\to+\infty$.
This problem was motivated by a question occuring in the theory of liquid crystals and appears also in superconductivity questions in large domains. 
\end{abstract}

\section{Introduction}
Let $\Omega\subset\R^3$ be an open bounded set with $\mj{C}^3$ boundary and $\A\in\mj{C}^2(\conj{\Om})$.\\
We will study the quadratic form $q_{\A}$ defined by :
$$q_{\A}(u)=\int_{\Om}\left|(i\nabla+q\A)u\right|^2 dx,\qquad \forall u\in H^1(\Om)$$
and consider the associated selfadjoint operator, i.e the Neumann realization of $(i\nabla+q\A)^2$ on $\Om$.
We denote $\mu_{\Om}(q,\A)$ or $\mu(q,\A)$ the lowest eigenvalue of the previous operator.
Our purpose is to study the behavior of this eigenvalue as $q$ tends to infinity and to control the uniformity of the estimates with respect to the magnetic field.\\
We let 
\bg{equation}\label{magneticfields}
\mj{A}=\{\A\in\mj{C}^3(\conj{\Om}) : |\B|=1 \mbox{ where } \B=\nabla\times\A\}.
\end{equation}
The main estimates obtained in this paper are summarized in the two following theorems :
\bg{theo}[Uniform lower bound]\label{lb}
For all $\ds{\eps\in]0,\frac{1}{2}[}$, there exists\\ 
\mbox { $C=C(\Om,\eps)>0$ } and $q_0>0$, such that, for all $q\geq q_0$ and for all $\A\in\mj{A}$,
\bg{equation*}
\mu(q,\A)\geq\Theta_0 q-C\left(q^{1-2\eps}+(1+|\nabla\B|_{\infty})q^{1/2+2\eps}\right).
\end{equation*}
\end{theo} 
\bg{theo}[Uniform upper bound]\label{ub}
For all $\delta\in]0,1/2[$, there exists $C=C(\Om,\delta)>0$ and $q_0=q_0(\Om;\delta)>0$, such that for all $q\geq q_0$ and all $\A\in\mj{A}$:
\bg{equation*}
\mu(q,\A)\leq\Theta_0 q+C(q^{2\delta}+|\B|^2_{\mj{C}^1}q^{2-4\delta}+|\B|_{\mj{C}^1}q^{1-\delta}+|\B|_{\mj{C}^2}q^{3/2-3\delta}+|\B|^2_{\mj{C}^2}q^{2-6\delta}),
\end{equation*}
where 
$|\B|^2_{\mj{C}^1}=|\B|_{\infty}+|\nabla\B|_{\infty}^2$ and $|\B|^2_{\mj{C}^2}=|\B|^2_{\mj{C}^1}+|\nabla^2\B|_{\infty}^2.$
\end{theo}
\bg{rem}
In those theorems, $\eps$ and $\delta$ are left undefined because they will play a role in the application to families of vector potentials where the semi-norms of $\B$ could become large. If the magnetic field is fixed and if we are not interested in uniformity, we take $\eps=\frac{1}{8}$ and $\delta=\frac{1}{3}$ to have the optimal estimates (relatively to the method), leading to the remainder $O(q^{3/4})$ in the first case (in fact, one can hope a remainder $O(q^{2/3})$) and $O(q^{2/3})$ in the second case (cf. \cite{HelMo2}). Let us also mention that, in dimension 2, the remainder is $O(q^{1/2})$ (see \cite{FouHel2}).
\end{rem}
Let us briefly recall the motivation of those estimates.
\paragraph{Liquid crystals}~\\
The first one occurs in the theory of liquid crystals.
The asymptotic properties of the Landau-de Gennes functional (cf. \cite{BCLP,Degennes,HelPan,Pan,Pan2}) lead to the analysis of the minimizers (or local minimizers) of the reduced functional : 
$$\mj{F}(\psi,\n)=\int_{\Om}|(i\nabla+q\n)\psi|^2+r|\psi|^2+\frac{g}{2}|\psi|^4 dx,$$
where $r>0$, $\psi\in H^1(\Om,\C)$ and $\n\in\mj{C}(\tau)$, with $\mj{C}(\tau)$ defined for $\tau>0$ by :
\bg{equation}\label{ctau}
\mj{C}(\tau)=\{Q\n_{\tau}Q^t, Q\in SO_3\},
\end{equation}
where $SO_3$ denotes the set of the rotations in $\R^3$ and
\bg{equation}\label{helic}
\n_{\tau}=(\cos(\tau x_3),\sin(\tau x_3),0).
\end{equation}
Let us notice that, if $\n\in\mj{C}(\tau)$, 
$$\nabla\times\n+\tau\n=0$$ 
and consequently 
$$|\nabla\times\n|_{\infty}=\tau.$$ 
Then, the analysis of the positivity of the Hessian of the functional at $\psi=0$ leads to a spectral problem and we are led to study the asymptotic properties of
\bg{equation}\label{mustar} 
\mu^*(q,\tau)=\inf_{\n\in\mj{C}(\tau)}\mu(q,\n),
\end{equation}
as $q\tau\to+\infty$.\\
In this context, X-B. Pan has given estimates (cf. \cite{Pan2}) as :
$$q\tau\to+\infty \quad\mbox{ and }\quad \tau\to 0$$
and Helffer and Pan give some extensions in \cite{HelPan} including the case :
$$q\tau\to+\infty \quad\mbox{ and }\quad \tau \mbox{ bounded}.$$
In this paper, we treat the case where :
$$q\tau\to+\infty \quad\mbox{ and }\quad \tau\to+\infty.$$
\paragraph{Superconductivity}~\\ 
The second one occurs in the theory of superconductivity in large domains.
In \cite{Almog2,Almog}, Y. Almog has analyzed properties of minimizers of Ginzburg-Landau's functional when the size of $\Om$ tends to infinity ; Theorems \ref{lb} and \ref{ub} permit to treat another regime (for the linear problem) ; in fact, $q$ will be allowed to tend to infinity.
\paragraph{Organization of the paper}~\\
The paper is organized as follows. 
First, we prove Theorem \ref{lb} in Section~2 and Theorem \ref{ub} in Section 3, then we will prove an Agmon estimate in Section 4 in order to study the localization of first eigenfunctions in the considered asymptotic regime. Finally, Section 5 will present the applications to the theory of liquid crystals and to the superconductivity in large domains. In each case we will show that the eigenfunctions become localized at the boundary. This corresponds to what is called surface smecticity in the first case and surface superconductivity in the second case (see \cite{Pan3}).

\section{Lower bound}
In this section, we give the proof of Theorem \ref{lb}. It is based on a localization technique through a partition of unity and the analysis of simplified models.
\subsection{Partition of unity}
For each $r>0$, we consider a partition of unity (cf. \cite{HelMo2}) with the property that there exists $C=C(\Om)>0$ such that :
\bg{eqnarray}\label{partition}
\sum_j |\chi_j^r|^2=1&\mbox{ on }\Om\, ;\\
\sum_j |\nabla\chi_j^r|^2\leq \frac{C}{r^2}&\mbox{ on }\Om.\\\nonumber
\end{eqnarray}
Each $\chi_j^r$ is a $\mj{C}^{\infty}$-cutoff function with support in the ball of center $x_j$ and radius $r$ (denoted by $B_j$). We will choose $r$ later for optimizing the error.
We will use the IMS formula (cf. \cite{Cycon}) :
\bg{lem}
\bg{eqnarray}\label{IMS}
q_{\A}(u)=\sum_j q_{\A}(\chi_j u)-\sum_j \||\nabla\chi_j^r|u\|^2,\qquad\forall u\in H^1(\Om).
\end{eqnarray}
\end{lem}
So, in order to minimize $q_{\A}(u)$, we will be reduced to the minimization of $q_{\A}(v)$, with $v$ supported in some $B_j$, the price to pay being an error of order $\frac{C}{r^2}$.
\subsection{Approximation by the constant magnetic field in a ball or a semi-ball}
We want to have estimates depending only on the magnetic field $\B=\nabla\times\A$, that's why we look for a canonical choice of $\A$ depending only on $\B$ ; it is the aim of the following lemmas.
Let $B$ a ball (or semi-ball) of center $0$ and radius $r>0$.
\bg{lem}
Let $\F\in\mj{C}^2(\conj{B},\R^3)$.\\
We assume the existence a constant $C>0$ such that :
$$|\nabla\times \F|\leq C|x|,$$
for $x\in\conj{B}$.
Then, there exists $u\in\mj{C}^3(\conj{B})$ and $\alpha>0$ a (universal) constant such that :
$$|\F(x)-\nabla u(x)|\leq \alpha C|x|^2,$$
for all $x\in\conj{B}$.
\end{lem}

\bg{preuve}
The proof is similar to the one of Poincaré's theorem.\\
Let us define, for all $x\in\conj{B}$ :
$$u(x)=\int_0^1 \F(tx)\cdot x\, dt.$$
Let us verify that $u$ is suitable. As $\F\in\mj{C}^2(\conj{B},\R^3)$, we can extend $\F$ in a $\mj{C}^2$ function on $\R^3$, so by computing, we have, for $x\in\conj{B}$ :
$$\dr_i u(x)=F_i(x)+\sum_{j=1,j\neq i}^3 \int_0^1 \left(\dr_i F_j-\dr_j F_i\right)(tx)t x_j dt.$$
\end{preuve}
We now state the lemma which will give us uniformity in our further estimates~:

\bg{lem}\label{jaugelocale}
There exists $C>0$ such that, for all $\A\in~\mj{C}^2(\overline{B})$, there exists $\phi\in\mj{C}^3(\conj{B})$ verifying :
$$|\A(x)-\A^{lin}(x)-\nabla \phi(x)|\leq C|\nabla\B|_{\infty}|x|^2,$$
for $x\in\conj{B}$ and where $\A^{lin}$ is defined by :
$$\A^{lin}(x)=\frac{1}{2}\B(0)\wedge x.$$
\end{lem}

\subsection{Local estimates for the lower bound} 
We now distinguish two cases : the balls inside $\Om$ and those which intersect the boundary. For the balls inside $\Om$, we are reduced to the problem of Dirichlet with constant magnetic field, and for the other balls, to the problem of Neumann on an half plane with constant magnetic field.
\subsubsection{Study inside $\Om$}
Let $j$ such that $B_j$ does not intersect the boundary.\\
Using Lemma \ref{jaugelocale} we make the change of gauge $v\mapsto e^{-i\phi}v$ and with the classical inequality :
$$|a+b|^2\geq (1-\lambda^2)|a|^2-\frac{1}{\lambda^2}|b|^2,$$
for $\lambda>0$, we get :
\bg{eqnarray*}
\int_{\Om}|(i\nabla+q(\A-\nabla\phi))(\chi_j u e^{-i\phi})|^2 dx&\geq& \Big((1-\lambda^2)\int_{\Om}|(i\nabla+q\A^{lin})(\chi_j u e^{-i\phi})|^2 dx\\
&&-C^2q^2\frac{|\nabla\B|^2_{\infty}}{\lambda^2}r^4\Big)\int_{\Om}|\chi_j u|^2 dx.
\end{eqnarray*}
We are reduced to the problem of Dirichlet with constant magnetic field and we have :
$$\int_{\Om}|(i\nabla+q\A^{lin})(\chi_j u e^{-i\phi})|^2 dx\geq q\int_{\Om}|\chi_j u|^2 dx.$$
Thus, we find :
$$\int_{\Om}|(i\nabla+q\A)\chi_j u|^2 dx\geq \left((1-\lambda^2)q-C^2q^2\frac{|\nabla\B|^2_{\infty}}{\lambda^2}r^4\right)\int_{\Om}|\chi_j u|^2 dx.$$
\subsubsection{Study near the boundary}
We refer to \cite{HelMo,HelMo2}, but we will control carefully the uniformity. We first recall some properties of the harmonic oscillator on an half axis (see \cite{DauHel,HelMo3}).
\paragraph{Harmonic oscillator on $\R_+$}~\\
For $\xi\in\R$, we consider the Neumann realization $\got{h}^{N,\xi}$ in $L^2(\R_+)$ associated with the operator \bg{equation}\label{oh}
-\frac{d^2}{dt^2}+(t+\xi)^2,\quad\mj{D}(\got{h}^{N,\xi})=\{u\in B^2(\R_+) : u'(0)=0\}.
\end{equation}
One knows that it has compact resolvent and its lowest eigenvalue is denoted $\mu(\xi)$ ; the associated $L^2$-normalized and positive eigenstate is denoted $u_{\xi}$ and is in the Schwartz class. The function $\xi\mapsto\mu(\xi)$ admits a unique minimum in $\xi=\xi_0$ and we let : $\Theta_0=\mu(\xi_0)$.\\
We now introduce local coordinates near the  boundary in order to compare with the harmonic oscillator on $\R_+$ :
\paragraph{Local coordinates near the boundary}~\\
Let's assume that $0\in\dr\Om$. In a neighborhood $V$ of $0$, we take local coordinates $(y_1,y_2)$ on $\dr\Om$ (via a $\mj{C}^3$ map $\phi$). We denote $\No(\phi(y_1,y_2))$ the interior unit normal to the boundary at the point $\phi(y_1,y_2)$ and define local coordinates in $V$ :
$$\Phi(y_1,y_2,y_3)=\phi(y_1,y_2)+y_3 \No(\phi(y_1,y_2)).$$
More precisely, for a point $x\in V$, $\phi(y_1,y_2)$ is the projection of $x$ on $\dr\Om\cap V$ and $y_3=d(x,\dr\Om)$.\\
Taking a convenient map $\phi$, we can assume 
$$\Phi(0)=0 \mbox { and } D_0\Phi=Id.$$
Let $j$ such that $B_j\cap\dr\Om\neq\emptyset$. We can assume that $x_j\in\dr\Om$ and $x_j=0$ without loss of generality. After a change of variables, we have :
$$\int_{\Om}|(i\nabla+q\A)\chi_j u|^2 dx=\int_{y_3>0}|(i\nabla_y+q\tilde{\A})\widetilde{\chi_j u}|_{(D\Phi)^{-1}((D\Phi)^{-1})^t}^2 |\det(D\Phi)| dy,$$
where the tilde denotes the functions in the new coordinates and 
$$\tilde{\A}=D_y\Phi(\A(\Phi(y))).$$
There exists $C>0$ (uniform in $j$) such that :
$$\int_{y_3>0}|(i\nabla_y+q\tilde{\A})\widetilde{\chi_j u}|_{(D\Phi)^{-1}((D\Phi)^{-1})^t}^2 |\det(D\Phi)| dy\geq (1-Cr)\int_{y_3>0}|(i\nabla_y+q\tilde{\A})\widetilde{\chi_j u}|^2 dy.$$
We use again the approximation by the constant magnetic field (for semi-balls) on the support of $\widetilde{\chi_j}$.\\
More precisely, there exists $\alpha>0$ uniform in $j$, such that $$\su(\widetilde{\chi_j})\subset B(x_j,\alpha r).$$
Then, we change our partition of unity : we replace the balls which intersect the boundary by $\Phi(B(x_j,\alpha r))$.
There exists $C>0$ such that for all $j$, there exists $\tilde{\A}^{lin}$ (defined in Lemma~\ref{jaugelocale}) satisfying :
\bg{eqnarray*}
\int_{y_3>0}|(i\nabla_y+q\tilde{\A})\widetilde{\chi_j u}|^2 dy&\geq& (1-\lambda^2)\int_{y_3>0}|(i\nabla_y+q\tilde{\A}^{lin})\widetilde{\chi_j u}|^2 dy\\
&&-\frac{C^2 q^2}{\lambda^2}|\nabla\tilde{\B}|^2_{\infty}r^4\int_{\Om}|\widetilde{\chi_j u}|^2 dy,
\end{eqnarray*}
where $\tilde{\B}=\nabla_y\times\tilde{\A}$. In order to express $\tilde{\B}$ as a function of $\B$, we need the following lemma :
\bg{lem}
With the previous notations, we have :
$$\tilde{\B}=\det(D\Phi)((D\Phi)^{-1})^t\B.$$
\end{lem}
\bg{preuve}
The result is standard. Let us recall the proof for completness. Let us introduce the 1-form $\omega$ :
$$\omega=A_1 dx_1+A_2 dx_2+A_3 dx_3.$$
In the new coordinates $x=\Phi(y)$, we have, with the previous notations :
$$\omega=\tilde{A}_1 dy_1+\tilde{A}_2 dy_2+\tilde{A}_3 dy_3.$$
Then, it remains to write :
$$d\omega=(\nabla\times\A)_1 dx_2\wedge dx_3+(\nabla\times\A)_2 dx_1\wedge dx_3+(\nabla\times\A)_3 dx_1\wedge dx_2,$$
and to express $dx_i$ as a function of $(dy_j)$.\\
The comatrix formula gives the conclusion.
\end{preuve} 
We are reduced to the case of constant magnetic field (of intensity $q$) on $\R^3_+=\{y_3>0\}$ (see \cite{HelMo,HelMo2,LuPan2}) and we get :
$$\int_{y_3>0}|(i\nabla_y+q\tilde{\A})\widetilde{\chi_j u}|^2 dy\geq \Theta_0 q\int_{y_3>0}|\widetilde{\chi_j u}|^2 dy.$$
Thus, we find :
$$\int_{y_3>0}|(i\nabla_y+q\tilde{\A})\widetilde{\chi_j u}|^2 dy\geq \left((1-\lambda^2)q\Theta_0-\frac{C^2 q^2}{\lambda^2}(1+|\nabla\B|^2_{\infty})r^4\right)\int_{y_3>0}|\widetilde{\chi_j u}|^2 dy.$$ 
\subsubsection{End of the proof}
We take, for $\ds{0<\eps<\frac{1}{2}}$, $r=\frac{1}{q^{1/2-\eps}}$. 
We divide by $q$ and we choose $\lambda$ such that :
$$\lambda^2=\frac{q}{\lambda^2}(1+|\nabla\B|^2_{\infty})r^4.$$
Then, the previous estimates lead to the existence of $C>0$ and $q_0>0$ depending only on $\Om$ such that for all $q\geq q_0$ :
$$\frac{q_{\A}(u)}{q}\geq \left(q\Theta_0-C\left(r+\lambda^2+\frac{1}{qr^2}\right)\right)\int_{\Om}|u|^2 dx.$$
We finally find, by the minimax principle :
$$\frac{\mu(q,\A)}{q}\geq\Theta_0-C\left(\frac{1}{q^{1/2-2\eps}}+\frac{1}{q^{2\eps}}+\frac{|\nabla\B|_{\infty}}{q^{1/2-2\eps}}\right).$$
\section{Upper bound}
In this section, we give a proof of Theorem \ref{ub}. We refer to \cite{HelMo2}.
In the case of the constant magnetic field on $\R_+^3$, we know (cf. \cite{HelMo,LuPan2}) that the bottom of the spectrum is minimal when the magnetic field is tangent to the boundary. So, we will look for a quasimode localized near a point where the magnetic field is tangent. Then, we will take as trial function some truncation of $u_{\xi_0}$.\\
So, we fix $x_0\in\dr\Om$ such that : $\B(x_0)\cdot\nu=0$. Such a $x_0$ exists ; indeed, noticing that $\dive(\B)=0$, the Stockes formula gives :
$$\int_{\dr\Om}\B\cdot\nu d\sigma=\int_{\Om}\dive(\B)dx=0.$$
We take $\delta\in]0,1/2[$ and we suppose that $u$ is such that $$\su(u)\subset B(x_0,\alpha r),$$
with $r=\frac{1}{q^{\delta}}$.\\
We assume that the support of $u$ is small enough and after a change of coordinates, we can use the same arguments as in Lemma \ref{jaugelocale} and take a gauge in which $\A$ satisfies :
$$|\A(y)-\A^0(y)|\leq C|\B|_{\mj{C}^2}|y|^3,$$
where $\A^0=\A^{lin}+R$, with $R=(R_1,R_2,R_3)$ and $R_j$ homogeneous polynomial of order 2 and where $\nabla^2\B$ denotes the hessian matrix of $\B$.\\
We find : 
$$q_{\A}(u)\leq (1+Cr)(q_{\A^0}(u)+C(|\B|^2_{\mj{C}^2}r^6 q^2\|u\|^2+qr^3|\B|_{\mj{C}^2}\|u\|q_{\A^0}(u)^{1/2})).$$
We let :
$$u(y)=q^{1/4+\delta}e^{-i\xi_0 y_2 q^{1/2}}u_{\xi_0}(q^{1/2}y_3)\chi(4q^{\delta}y_3)\chi(4q^{\delta}(y_1^2+y_2^2)^{1/2}).$$
We have to compare : $q_{\A^0}(u)$ and $q_{\A^{lin}}(u)$.
We get : 
\bg{eqnarray*}
q_{\A^0}(u)&\leq &q_{\A^{lin}}(u)+Cq^2 r^4(1+|\nabla\B|^2_{\infty})\|u\|^2\\
&&+2\Re\left\{\int_{|y|\leq\alpha r, y_3>0}(i\nabla+q\A^{lin})u\cdot (q\A^0-q\A^{lin})u dy\right\}.
\end{eqnarray*}
We have to estimate the double product (cf. \cite[section 6,p. 120]{HelMo2}) : 
$$\left|\Re\left\{\int_{|y|\leq\alpha r, y_3>0}(i\nabla+q\A^{lin})u\cdot (q\A^0-q\A^{lin})u dy\right\}\right|\leq C(1+|\nabla\B|_{\infty})q^{1-\delta}.$$
Moreover, using the exponential decrease of $u_{\xi_0}$, we have : 
$$q_{\A^{lin}}(u)\leq \Theta_0 q+Cq^{2\delta}.$$
\section{Agmon's estimates}
In order to estimate the asymptotic localization of the first eigenfunctions in the applications, we will need Agmon's estimates to have some exponential decrease inside $\Om$ ; that is the aim of this section.
Let us introduce some notations (see \cite{Agmon, Almog, FouHel2}).
For $\gamma>0$ small enough, let $\eta_{\gamma}$ be a smooth cutoff function such that :
$$\eta_{\gamma}=\left\{
\bg{array}{cc}
1& \mbox{ if } d(y)=d(y,\dr\Om)\geq \gamma\\
0& \mbox{ if } y\notin\Om\\
\end{array}\right.,$$
with
$$|\nabla\eta_{\gamma}|\leq \frac{C}{\gamma}.$$
We let : $\Om_{\gamma}=\{y\in\Om : d(y,\dr\Om)\geq\gamma\}$.
For $\alpha>0$, we let $$\xi(y)=\eta_{\gamma} e^{\alpha d(y)}.$$
We finally denote $\mu_0(q,\A)$ the lowest eigenvalue of the Dirichlet's realization of $(i\nabla+q\A)^2$ on $\Om$.
We have the following localization property :
\bg{prop}\label{agmon}
There exists $C>0$ and $\gamma_0>0$, depending only on $\Om$ such that for all $0<\eps\leq 1$ and $\alpha$ verifying :
$$0<\alpha<\left(\frac{1}{1+\eps}\right)^{1/2}(\mu_0-\mu)^{1/2},$$
and for all $0<\gamma\leq\gamma_0$, if $u$ is a normalized mode associated with the lowest eigenvalue $\mu=\mu(q,\A)$ of the Neumann realization of $(i\nabla+q\A)^2$, then :
$$\|\eta_{\gamma}e^{\alpha d(y)}|u|\|_{H^1(\Om)}\leq \frac{C}{\sqrt{\eps\gamma}}\left(\frac{\mu_0+1}{\mu_0-\mu-(1+\eps)\alpha^2}\right)^{1/2}e^{\alpha\gamma}.$$
\end{prop}
\bg{preuve}
We consider the equation verified by $u$ : $$(i\nabla+q\A)^2 u=\mu u.$$
One multiplies by $\xi^2 \conj{u}$ and integrate by parts (using $(i\nabla+q\A)u\cdot\nu=0$ on $\dr\Om$) to get :
$$|(i\nabla+q\A)(\xi u)|_2^2=\mu|\xi u|_2^2+|(\nabla\xi) u|_2^2.$$
We have, for all $\eps>0$ : 
$$|(\nabla\xi)u|^2\leq (1+\frac{1}{\eps})\int_{\Om\setminus\Om_{\gamma}}|\nabla\eta|^2e^{2\alpha d(y)}|u|^2+(1+\eps)\alpha^2\int_{\Om}|\xi u|^2.$$
We use that $|u|_2=1$ to find :
$$|\xi u|^2\leq \frac{C}{\gamma}(1+\frac{1}{\eps})\frac{e^{2\alpha\gamma}}{\mu_0-\mu-(1+\eps)\alpha^2}.$$
Moreover, the diamagnetic inequality gives :
$$|\nabla|\xi u||^2\leq |(i\nabla+q\A)(\xi u)|^2.$$
It follows that :
$$||\xi|u|||^2_{H^1(\Om)}\leq \frac{C}{\gamma}(1+\frac{1}{\eps})e^{2\alpha\gamma}\frac{\mu_0+1}{\mu_0-\mu-(1+\eps)\alpha^2}.$$
\end{preuve}
\section{Applications}
We now describe two applications of our main results.
\subsection{Application to an helical vector field}
In this section, we study $\mu^*(q,\tau)$ defined in (\ref{mustar}) as $q\tau\to+\infty$ and $\tau\to+\infty$. Due to the definition (\ref{mustar}), we will use the uniform analysis of $\mu(q,\n)$ with $\n\in\mj{C}(\tau)$.
\subsubsection{Estimate of the first eigenvalue}
The main theorem in this section is the following :
\bg{theo}\label{estimatepan}
Let $c_0>0$ and $0\leq x<\frac{1}{2}$. There exists $C>0$ and $q_0>0$ depending only on $\Om$, $c_0$ and $x$ such that, if $(q,\tau)$ verifies $q\tau\geq q_0$ and 
\bg{equation}\label{tau}
\tau\leq c_0 (q\tau)^x,
\end{equation}
then :
\bg{equation}\label{estimatepan2}
\Theta_0-\frac{C}{(q\tau)^{1/4-x/2}}\leq\frac{\mu^*(q,\tau)}{q\tau}\leq\Theta_0+\frac{C}{(q\tau)^{1/3-2x/3}}.
\end{equation}
\end{theo}
\bg{rem}
This statement was obtained for $x=0$ in \cite{HelPan} and rough estimates where given in \cite{BCLP} as $\ds{\frac{\tau}{q}\to 0}$.
\end{rem}
\bg{preuve}
Let us notice that : $$\mu(q\tau,\frac{Q\n_{\tau}Q^t}{\tau})=\mu(q,Q\n_{\tau}Q^t).$$
Moreover, there exists $C>0$ such that for all $\tau>0$ and $\n\in\mj{C}(\tau)$, if $\ds{\A=\frac{\n}{\tau}}$ and $\B=\nabla\times\A$, then : 
$$|\B|_{\infty}=1,$$
$$|\nabla\B|_{\infty}\leq C\tau,$$
$$|\nabla^2\B|_{\infty}\leq C\tau^2.$$
For the lower bound, we apply Theorem \ref{lb} to the subfamily $\mj{C}(\tau)$ of $\mj{A}$ and, using (\ref{tau}), we get :
$$\frac{\mu^*(q,\tau)}{q\tau}\geq \Theta_0-C\left(\frac{1}{(q\tau)^{1/2-2\eps}}+\frac{1}{(q\tau)^{2\eps}}+c_0\frac{(q\tau)^x}{(q\tau)^{1/2-2\eps}}\right).$$ 
We choose $\eps$ such that $\frac{1}{2}-2\eps-x=2\eps$, i.e : $2\eps=\frac{1}{4}-\frac{x}{2}$.\\
For the upper bound, we apply Theorem \ref{ub} with $\ds{\A=\frac{\n_{\tau}}{\tau}}$.
Then, we get :
$$\mu^*(q,\tau)\leq \Theta_0 q\tau+C((q\tau)^{2\delta}+(q\tau)^{2-4\delta+2x}+(q\tau)^{1-\delta+x}+(q\tau)^{3/2-3\delta+2x}+(q\tau)^{2-6\delta+4x}).$$
We choose $\delta$ such that : $$2\delta=1-\delta+x.$$
Thus, we take $\delta=\frac{1+x}{3}$ and the upper bound follows.
\end{preuve}

\subsubsection{Localization of the ground state near the boundary as $\tau~\to~+~\infty$}
We first state a proposition :
\bg{prop}
For $\A\in\mj{A}$ (cf. (\ref{magneticfields})), we denote $\mu_0(q,\A)$ the bottom of the spectrum of the Dirichlet realization of $(i\nabla+q\A)^2$.\\ 
Then, for all $\eps\in]0,1/2[$, there exists $C=C(\Om,\eps)>~0$ and $q_0=q_0(\Om,\eps)$ s.t. if $q\geq q_0$, for all $\A\in\mj{A}$ : 
$$\frac{\mu_0(q,\A)}{q}\geq 1-C\left(\frac{1}{q^{2\eps}}+\frac{|\nabla\B|_{\infty}}{q^{1/2-2\eps}}\right).$$
\end{prop}
\bg{preuve}
We again use partition (\ref{partition}), formula (\ref{IMS}) and the proof is the same as for Theorem \ref{lb}. 
\end{preuve}
We deduce :
\bg{cor}
Let $c_0>0$.
For all $x\in[0,1/2[$, there exists $C=C(\Om,x,c_0)>~0$, such that, for all $\n\in\mj{C}(\tau)$ and $(q,\tau)$ such that $\tau\leq c_0 (q\tau)^x$ :
$$\frac{\mu_0(q\tau,\frac{\n}{\tau})}{q\tau}\geq 1-C\left(\frac{1}{q\tau}\right)^{1/4-x/2}.$$
\end{cor}
As an immediate consequence of Proposition \ref{agmon}, we have the following theorem :
\bg{theo}\label{agmontau}
For all $x\in[0,1/2[$, there exists $\delta_0>0$, $C>0$, $c>0$ such that if $(q,\tau)$ verifies $q\tau\geq\delta_0$ and $\tau\leq c_0 (q\tau)^x$, then
for all $\n\in\mj{C}(\tau)$ and $u$ a $L^2$-normalized solution of
$$
\bg{array}{c}
(i\nabla+q\n)^2u=\mu(q,\n)u,\quad\mbox{ in $\Om$ }\\
(i\nabla+q\n)u\cdot\nu =0,\quad\mbox{ on $\dr\Om$}
\end{array}
$$
we have :
$$\|\eta_{\frac{c}{\sqrt{q\tau}}}e^{((1-\Theta_0)^{1/2}\sqrt{q\tau}-r(q\tau))d(\cdot,\dr\Om)}|u|\|_{H^1(\Om)}\leq C,$$
where $r(q\tau)=(q\tau)^{3/8+x/4}.$
\end{theo}
\bg{preuve}
We apply Proposition \ref{agmon} with $\eps=(q\tau)^{-1/4+x/2},$ $\alpha=(1-\Theta_0)^{1/2}(\sqrt{q\tau}-\lambda),$ $\gamma=\frac{1}{\sqrt{q\tau}},$ $\lambda=(q\tau)^{1/4+x_1/2}, \mbox{ with } x<x_1<\frac{1}{2}$ 
and we notice, by taking a truncated gaussian, that 
$$\mu_0(q\tau,\frac{\n}{\tau})\leq C q\tau.$$
We finally take $x_1=\frac{1}{2}(\frac{1}{2}+x)$ and apply the estimate (\ref{estimatepan2}) of Theorem \ref{estimatepan}.
\end{preuve}

\subsection{Surface superconductivity in large domains}
\subsubsection{The problem}
Let $R>0$ and $x_0\in\Om$. We denote $\Om_R=\{x_0+R(x-x_0),\,\, x\in\Om\}$. 
The aim of this part is to study the behaviour of $\mu_{\Om_R}(q,\A)$, where $\A\in\mj{A}$ (cf.(\ref{magneticfields})) as $q\to+\infty$ and $R\to+\infty$.
In his work \cite{Almog}, Almog studies the regime $q$ fixed and $R\to+\infty$ (see Theorem 1.1 in \cite{Almog}) and then makes $q$ to tend to infinity (see Section 3 of \cite{Almog}). 
In this section, we give a theorem which treats another regime~: $q\to+\infty$ and $R$ with polynomial increase in $q$.
We first observe the following scaling invariance :
\bg{lem}\label{rescaling}
Let $R>0$. We have :
$$\mu_{\Om_R}(q,\A)=\frac{1}{R^2}\mu_{\Om}\left(qR^2,\frac{\A(R\cdot)}{R}\right).$$
\end{lem}
\subsubsection{Estimate of the lowest eigenvalue}
\bg{theo}\label{estimatealmog}
Let $c_0>0$ and $y\geq 0$. There exists C>0, $q_0>0$ and $R_0>0$ depending only on $\Om$, $c_0$ and $y$ such that, if $(q,R)$ satisfies $q\geq q_0$, $R\geq R_0$ and $\ds{R\leq c_0 q^y}$, then, for $\A\in\mj{A}$ :
$$\Theta_0 -\frac{C}{(qR^2)^{\frac{1}{4(1+2y)}}}\leq\frac{\mu_{\Om_R}(q,\A)}{q}\leq\Theta_0-\frac{C}{(qR^2)^{\frac{1}{3(1+2y)}}}.$$ 
\end{theo}

\bg{preuve}
We use Lemma \ref{rescaling}.
We let $\ds{x=\frac{y}{1+2y}}$ and notice that $R\leq c(y)(qR^2)^x$ with $\ds{c(y)=c_0^{\frac{1}{1+2y}}}$ and $\frac{\A(R\cdot)}{R}\in\mj{A}$.
Then by the Theorems \ref{lb} and \ref{ub}, we have the wished conclusion by using the same arguments as for Theorem \ref{estimatepan}.
\end{preuve}

\subsubsection{Localization of the groundstate near the boundary in large domains}
In the case of large domains, we prove a quite analogous theorem with Theorem \ref{agmontau} :
\bg{theo}
For all $y\geq 0$, there exists $\delta_0>0$, $\delta_1>0$, $C>0$, $c>0$ such that if $(q,R)$ verifies $q\geq\delta_0$, $R\geq\delta_1$ and $R\leq c_0 q^y$, then
for all $\A\in\mj{A}$ and $u$ a $L^2$-normalized solution of
$$
\bg{array}{c}
(i\nabla+q\A)^2u=\mu_{\Om_R}(q,\A)u,\quad\mbox{ in $\Om_R$ }\\
(i\nabla+q\A)u\cdot\nu =0,\quad\mbox{ on $\dr\Om_R$}
\end{array}
$$
we have,
$$\|\eta_{\frac{c}{\sqrt{q}}}e^{(1-\Theta_0)^{1/2}(\sqrt{q}-r(q,R))d(\cdot,\dr\Om_R)}|u|\|_{H^1(\Om_R)}\leq C,$$
where $r(q,R)=q^{1/2-\frac{1}{8(1+2y)}}R^{-\frac{1}{4(1+2y)}}.$
\end{theo}
\bg{preuve}
After a rescaling, the proof is the same as the one of Theorem \ref{agmontau}. 
\end{preuve}

\addcontentsline{toc}{section}{References}
\bibliographystyle{alpha}
\bibliography{biblio}
\end{document}